\documentclass[aps,twocolumn,pra,showpacs,amsmath,amssymb,longbibliography,showkeys,10pt]{revtex4}
\usepackage{graphicx}
\usepackage{epstopdf}
\usepackage[normalem]{ulem}
\usepackage{braket}
\usepackage{hyperref}
\usepackage[dvipsnames]{xcolor}
\usepackage{ulem}
\allowdisplaybreaks

\begin{document}

\title{Decoherence of Schr\"odinger cat states in light of wave/particle duality}

\author{Th. K. Mavrogordatos}
\email{themis.mavrogordatos@fysik.su.se}
 \affiliation{Department of Physics, AlbaNova University Center, SE 106 91, Stockholm, Sweden}

\date{\today}

\begin{abstract}
We challenge the standard picture of decohering Schr\"odinger cat states as an ensemble average obeying a Lindblad master equation, brought about locally from an irreversible interaction with an environment. We generate self-consistent collections of pure system states correlated with specific environmental records, corresponding to the function of the wave-particle correlator first introduced in Carmichael {\it et al.} [Phys. Rev. Lett. 85, 1855 (2000)]. In the spirit of Carmichael {\it et al.} [Coherent States: Past, Present and Future, pp. 75--91, World Scientific (1994)], we find that the complementary unravelings evince a pronounced disparity when the ``position'' and ``momentum'' of the damped cavity mode---an explicitly open quantum system---are measured. Intensity-field correlations may largely deviate from a monotonic decay, while Wigner functions of the cavity state display contrasting manifestations of quantum interference when conditioned on photon counts sampling a continuous photocurrent. In turn, the conditional photodetection events mark the contextual diffusion of both the net charge generated at the homodyne detector, and the electromagnetic field amplitude in the resonator.   
\end{abstract}


\maketitle

\section{Introduction}

Coherent states occupy a central position in quantum electrodynamics (QED). They create a connection between quantum and semiclassical theories of photoelectric detection~\cite{Mandel1958,KelleyKleiner1964,Davidson1968,Kimble1977,Saleh1978,SrinivasDavies1981,Mandel1981,SrinivasDavies1982,Carmichael1993QTI}: being eigenstates of an operator that annihilates photons from the electromagnetic field, they are natural candidates of quantum states of light that have the same effect on a photoelectric detector as coherent fields. Coherent states are also essential in telling apart classical and non-classical optical fields, featuring in the definition of the Glauber--Sudarshan $P$-representation~\cite{Sudarshan1963, Glauber1963Rep} which sets the boundary. In fact, Glauber established their central place in his early work on quantum theory of coherence, which revolved around an analysis of photoelectric detection~\cite{Glauber1963PC, Glauber1963Coh, Glauber1963Rep}. With the advent of cavity and circuit QED, non-classical states of light were routinely within experimental reach and control in configurations where one atom, be it natural or artificial strongly interacts with one or a few photons. In situations of the like, the $P$-representation no longer maps quantum dynamics into a classical stochastic process, while proposed modifications to press on with such a mapping come with their own shortcomings~\cite{CarmichaelBook2}. 

Keeping the master equation (ME) description of a single decaying cavity mode as an open QED system explicitly in mind~\cite{Davies1976,Gerry1997, BreuerPetruccione,Carmichael2013Ch4,Minganti2016,Mamaev2018,Lebreuilly2019,Zhou2021,Zapletal2022,Krauss2023,Kozin2024}, the formalism of quantum trajectories starts with photoelectric detection and addresses the following question: How does the evolution of the quantum oscillator state run in parallel with the classical stochastic process of photoelectric counts? The answer is given by a quantum mechanical theory which is able to simulate the evolution of the oscillator before taking the ensemble average to form the reduced system density operator $\rho(t)$. In this process, a quantum and a classical stochastic process are consistently coupled. Here we will make use of this coupling to investigate the decay of macroscopic superposition states~\cite{Walls1985,Phoenix1990,Kim1992,Brune1992,Zurek2003,RomeroIsart2010,Carmichael2013Ch4, Girvin2019,Dakic2017,Qin2019,Qin2021} in conditional homodyne detection~\cite{YurkeStoler1987,Schleich1991,Carmichael2000, CarmichaelFosterChapter, Carmichael2004,MarquinaCruz2008}, an extension of the intensity correlation technique and its reliance on a {\it conditional} measurement, introduced by Hanbury-Brown and Twiss~\cite{Brown1956, BrownHTwissI, BrownHTwissII}. In 1986, Yurke and Stoller~\cite{YurkeStoler1986} proposed an idea on how a macroscopic superposition state might be prepared and subsequently observed by means of homodyne detection~\cite{Wiseman1993, Carmichael1993QTIII, CarmichaelBook2,wiseman_milburn_2009}. Several alternative schemes and physical systems have been suggested since~\cite{Wolinsky1988,Song1990,Brune1992,Monroe1996,Dakna1997,Agarwal1997,Gerry1999, Lund2004,Jeong2005,HarocheBook,Glancy2008,Vlastakis2013,Haroche2013,Leghtas2015,Bergmann2016,Michael2016,Ofek2016,Liao2016,Wang2016, Girvin2019,Song2019,Omran2019,Joshi2021,Lewenstein2021,Rivera-Dean2021,Rivera-Dean2022,Cosacchi2021,Pogorelov2021,Takase2021,Zhou2021,Wang2022,He2023,Ayyash2024,Bocini2024,Kozin2024,
Hotter2024,Yu2024,Torres2024}, while later work also established that the photoelectron counting distribution in homodyne detection is given by a marginal of the Wigner function representing the state of the cavity---the local oscillator phase determines the marginal~\cite{Vogel1989, Smithey1993}. 

\begin{figure*}
\includegraphics[width=0.72\textwidth]{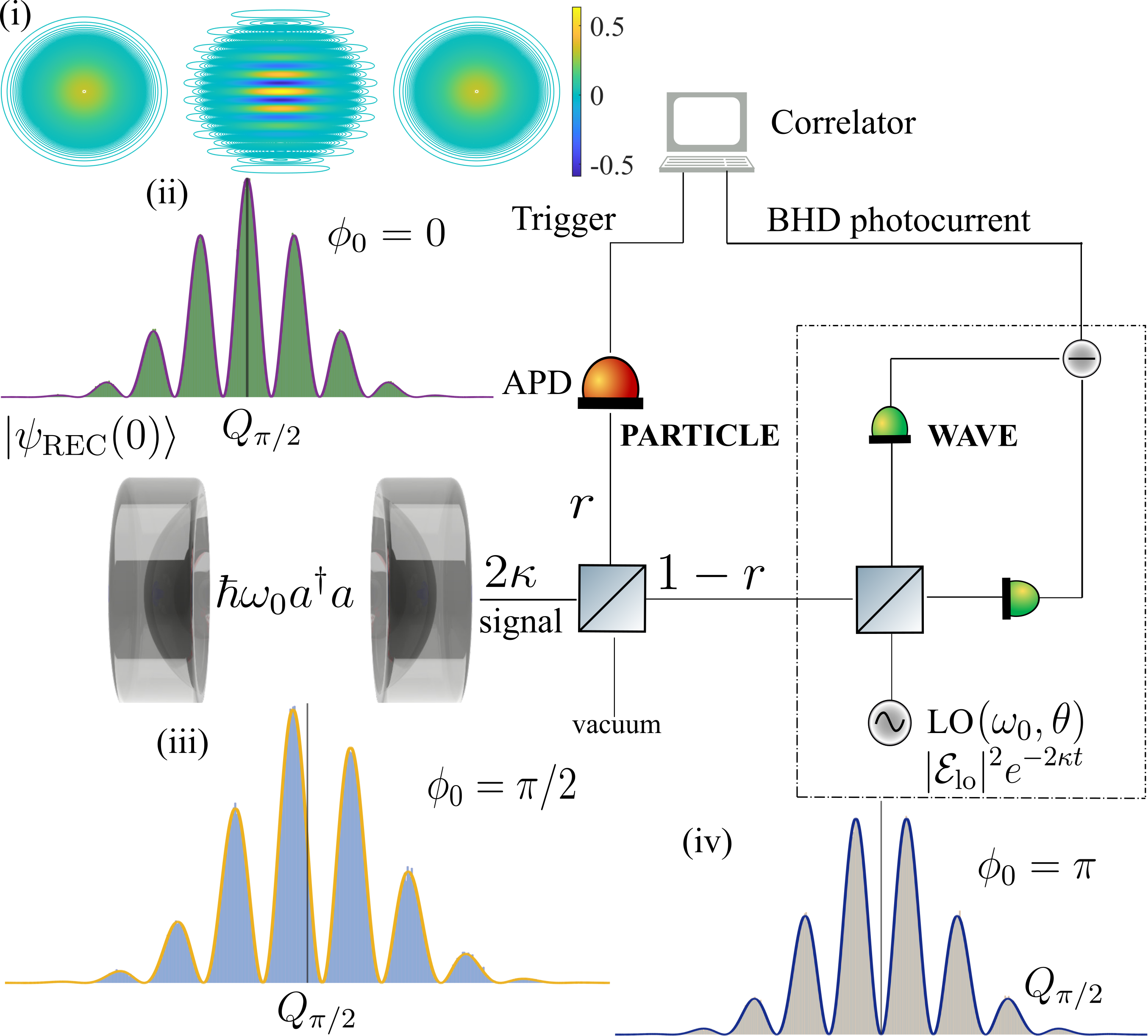}
\caption{Schematic illustration of the wave-particle correlator (center) realizing conditional homodyne detection, with a macroscopic coherent superposition as an initial {\it source state}. The device operates with a fraction $r$ of the input light flux going to an avalanche photodiode (APD) in the ``trigger'' channel, and the remaining $1-r$ directed to a balanced mode-matched homodyne detector (BHD). The BHD samples the quadrature phase amplitude that is in phase with the local oscillator (LO) field; $\theta$ is the LO phase. The cavity mode $a$ is prepared in the superposition of coherent states~\eqref{eq:incat} at time $t=0$, from which it decays to produce the scattered field (signal). Inset {\bf (i)} depicts a contour plot of the Wigner function $W(x+iy;t=0)$ of the initial state~\eqref{eq:incat}, with $A=4$ and $\phi_0=0$. Insets {\bf (ii)--(iv)} depict histograms of the cumulative charge $Q_{\theta=\pi/2}$ deposited in the BHD after the light has left the cavity, when the correlator operates with $r=0$ and $\theta=\pi/2$, for an initial Schr\"odinger cat state with $A=4$ and: $\phi_0=0$ {\bf (ii)}, $\pi/2$ {\bf (iii)} and $\pi$ {\bf (iv)}. The vertical lines indicate the location of $Q_{\pi/2}=0$.}
\label{fig:FIG1}
\end{figure*}

Data of the discrete, particle type, and continuous wave type are simultaneously collected~\cite{Carmichael2001}, such that light scattered from a cavity initially prepared in a Schr\"odinger cat state~\cite{Schrodinger1935, Dodonov1974, Gerry1997, HarocheBook, Nielsen2006, Ourjoumtsev2006, Ourjoumtsev2007, Sychev2017, Hacker2019,Pan2023} is seen in the simulated experiment to act as particle and wave. Both attributes serve to explain why $\rho(t)$ changes from a pure-state to a mixed-state density operator in a time much shorter than the cavity decay time by means of an unbalance in the two components of the superposition, operationally ascertained. Furthermore, the data correlate the quantum interference of a macroscopic Schr\"odinger cat with the emission of cavity photons. We will find that while these photon emissions sample the quadrature amplitude, the interference fringes in phase space conditionally resolve an accumulated amplitude and phase diffusion. At the two ends of the accomplished wave-particle correlator unravelings sit direct detection and (balanced) homodyne detection, exclusively pertaining to the corpuscular and wave attributes of light, respectively. Complementary displays of steady-state bimodality for a single nonlinear Kerr resonator with two photon driving depend on the measurement protocol, as reported in~\cite{Bartolo2017}. The bistable switching occurs either between even and odd cat states under direct photodetection, or between the coherent states defining a statistical mixture for high intracavity excitation, under homodyne detection. 

Coupling superposition states to another subsystem readily tracks the operational consequences of quantum coherence. For instance, a joint measurement of a shifted parity operator~\cite{Lutterbach1997, Girvin2019} and the projection of the Bloch vector of an atom entangled to the cat state leads to a correlation where the Wigner function of the cavity is weighted by the atomic spin orientations~\cite{Wodkiewicz2000}. In a twist, detecting dipole radiation from the dressed states of Jaynes--Cummings interaction~\cite{Carmichael1993QTIV,Alsing1991,Alsing1992,Solano2003} in a phase sensitive way via homodyne detection realizes an optical analogue of the Stern--Gerlach experiment~\cite{Venugopalan1995, Venugopalan1997} where the conditioned wavefunction makes a selection between initially superposed dressed states~\cite{CarmichaelSG1994}. More recently, in a dispersive detection circuit QED setup, a combination of heterodyne and homodyne detection was used for a superconducting qubit subject to decoherence owing to both relaxation and dephasing, attaining a full quantum-state tomography using incompatible and simultaneous measurements~\cite{Qubit2018}. Meanwhile, a ``single quadrature'' measurement accomplished by coherently driving a qubit and modulating cavity sidebands with a relative phase produced an angular diffusion akin to a random walk~\cite{Qubit2016}; in both cases, the dynamical evolution of the qubit state was modelled  by a stochastic ME~\cite{Qubit2016, Qubit2018}.

Conditional homodyne detection of a single system mode prepared in a coherent-state superposition resolves correlations similar to those read from entangled subsystems in various configurations~\cite{Fortunato2000,Li2017,Mamaev2018,Lebreuilly2019,DiGiulio2019,Kfir2019,Dahan2021,Cosacchi2021,Qin2019,
Stammer2022,Wang2022,Zapletal2022,Pan2023,Torres2024}. It does so by preserving or destroying the quantum coherence monitored through continuous measurement~\cite{Davies1976,Bergquist1986,Nagourney1986,Sauter1986,Cook1988,Belavkin1990,Barchielli1991,Dalibard1992,Gardiner1992,Dum1992, Molmer1993, Blanchard1995, gisin1997quantum, Plenio1998,Percival1998,Mabuchi1998,Peil1999,Brun2002,Wiseman2002,Gleyzes2007,Hegerfeldt2009, BarchielliGregoratti2012,Minganti2016,Minev2019} simulating individual experimental runs where the output field is detected.

The central aim of this report is to compare and connect key complementary methods of record making to manifestations of the coherence characterizing the free decay of an initial macroscopic state superposition. The attribute {\it operational}, frequently employed in this report, refers to the photon-counting sequences and/or deposited charge sets measured in quantum optical experiments, within the framework of photon scattering theory. In Sec.~\ref{sec:MEDD}, we introduce the simplest possible formulation to account for the interaction between system and reservoir on an ensemble average level. We also recall the role of decoherence in destroying quantum interference through the illustrative example of direct detection. Section~\ref{sec:CompUn} takes us to an unraveling strategy of the master equation accomplished by a device which underscores the subtlety involved in the coexistence of waves and particles under Bohr's complementarity. A particular operation mode of such device links the statistical behaviour of the charge produced by a homodyne detector to a phase difference between the two components of a macroscopic state superposition, as discussed in Sec~\ref{sec:chdist}. A transient and conditioned reading of coherence along single realizations is examined in Sec~\ref{sec:condintf}, where the difference between probing ``position'' and ``momentum''. The rapid decoherence time is contrasted to the time required for the interference pattern to take shape in the dynamically unfolding distribution of the accumulated charge as an ensemble average. The last part of our discussion in Sec.~\ref{sec:photonemdiff} is devoted to yet another distribution of the charge produced on one arm of the correlator, this time one which is conditioned on photon clicks at the other arm. Concluding remarks close the report out. 

\section{Master equation for free decay, and direct detection}
\label{sec:MEDD}

We are concerned with a set of quantum-trajectory unravelings of the primordial Lindblad master equation~\cite{Lindblad76} modelling the decay of a single cavity mode with frequency $\omega_0$ to a surrounding reservoir in the vacuum state~\cite{CarmichaelBook1, BreuerPetruccione}:
\begin{equation}\label{eq:ME}
\frac{d\rho}{dt}=\kappa (2a \rho a^{\dagger}-a^{\dagger}a \rho - \rho a^{\dagger}a),
\end{equation}
written in the interaction picture with respect to the system Hamiltonian $H_S=\hbar \omega_0 a^{\dagger}a$; here, $2\kappa$ is the photon loss rate. At $t=0$, the cavity mode $a$ is prepared in the macroscopic superposition state
\begin{equation}\label{eq:incat}
|\psi_{\rm REC}(0)\rangle=\frac{|A\rangle + e^{i\phi_0}|-A\rangle}{\sqrt{2[1+\cos\phi_0\exp(-2A^2)]}},
\end{equation}
a Schr\"odinger cat state with a fixed phase difference between its two components. In the above, $|A\rangle$ is a coherent state and $A$ is any positive number, while  $0\leq \phi_0<2\pi$. Even and odd cat states have $\phi_0=0$ and  $\pi$, respectively.

A coherent-state superposition of the form~\eqref{eq:incat} provides the canonical illustration of the role of decoherence in the rapid destruction of quantum interference~\cite{Brune1996, Deleglise2008}, much faster than the energy decay time $(2\kappa)^{-1}$. Under direct detection, the measurement scheme suggested by the very form of ME~\eqref{eq:ME}, the evolution of the conditioned state consists of a sequence of jumps, with continuous evolution in between. The jumps occur at the ordered count times $t_1, t_2, \ldots t_n$ over an interval of length $t$, at which the photoelectron ``clicks'' are registered by detectors placed in the environment surrounding the cavity. The un-normalized conditioned state $|\overline{\psi}_{\rm REC}\rangle$ keeps track of the emission sequence, putting together discontinuous jumps and coherent evolution~\cite{Carmichael2013Ch4}:
\begin{equation}\label{eq:psiDP}
\begin{aligned}
&|\overline{\psi}_{\rm REC}(t)\rangle=(\sqrt{2\kappa} A\,e^{-\kappa t_n})\ldots (\sqrt{2\kappa} A\,e^{-\kappa t_1})\\
&\times  \exp\left[\tfrac{1}{2} A^2 (1-e^{-2\kappa t})\right] \frac{|A e^{-\kappa t} \rangle + (-1)^n e^{i\phi_0} |-A e^{-\kappa t }\rangle}{\sqrt{2[1+\cos\phi_0\exp(-2A^2)]}},
\end{aligned}
\end{equation}
from which we obtain the record probability density as $\langle \overline{\psi}_{\rm REC}(t)|\overline{\psi}_{\rm REC}(t) \rangle$. The term $\exp\left[\tfrac{1}{2} A^2 (1-e^{-2\kappa t})\right]$ stems from combining all the pieces of continuous evolution between the jumps, and equals the probability of no photoelectron counts in the interval considered---the so-called null-measurement probability which stands out as a non-trivial prediction of quantum trajectory theory. The probability of the null-measurement result sequence is multiplied by the probability density $2\kappa A^2 e^{-2\kappa t_k}$ for individual counts at the times $t_k$, with $k=1,2,\ldots n$. 

For an ensemble of such realizations the number of photon emissions up to a given time cannot be predicted. Upon a lapse of a time interval $t\sim (2\kappa A^2)^{-1}$, the time waited for the escape of the very first photon from the cavity, any ensemble will contain an equal number of sequences with $n$ even and with $n$ odd for $A \gg 1$. Consequently, the interference fringes will be cancelled in the ensemble average in a very short time. The form of Eq.~\eqref{eq:psiDP} suggests that maintaining the coherence between the two components of the superposition boils down to tracking every single photon leaving the cavity and transferred to the environment. This instance creates a ``connection'' between system and environment: we are not to strictly pronounce the cat ``dead'' in view of the rapid decoherence. Rather, if we are to retain access to the coherence of the prepared cat state, we need to know what part of it is on the inside and what on the outside of the cavity {\it down to the level of one photon}~\cite{Carmichael2013Ch4}. 

\section{Complementary unravelings for wave/particle duality}
\label{sec:CompUn}

Direct detection is only one of the infinite possible methods of record making. In this part, we explore complementary unravelings~\cite{Carmichael1993QTII, Carmichael1999,Wiseman2012} produced under the action of the wave-particle correlator~\cite{Carmichael2000, Foster2000,Reiner2001,CarmichaelFosterChapter} in the following fashion. Photons (particles) trigger ``clicks'' in an avalanche photodiode (APD) resetting conditioned records of an electromagnetic field amplitude (wave) in the photocurrent output of a balanced homodyne detector (BHD)~\cite{Yuen1983}.

\subsection{Conditioned state and sampling process}

The BHD samples the quadrature phase amplitude with the local oscillator field phase $\theta$ (for $0\leq \theta <\pi$), defined as the operator $2\sqrt{2\kappa (1-r)}\,A_{\theta}$, where
\begin{equation}
A_{\theta} \equiv \tfrac{1}{2}(a e^{-i\theta}+a^{\dagger} e^{-i\theta})
\end{equation}
and $0\leq r \leq 1$. Meanwhile, the local-oscillator photon flux $|\mathcal{E}_{\rm lo}|^2 e^{-2\kappa t}$ (with $|\mathcal{E}_{\rm lo}|^2 \gg \kappa A^2$) is matched to the decaying signal flux, to perform what is termed a {\it mode matched} conditional homodyne detection. The charge $dq_{\theta}$ deposited in the detector circuit~\footnote{The term {\it charge} stands for a broadly defined detected signal corresponding to the generalized detector gain $G$, in view of the current circuit QED architectures in the microwave regime.}  in the interval from $t$ to $t+dt$ generates the BHD photocurrent $I_{\theta}(t)$ via $dI_{\theta}=-\tau_{d}^{-1}(I_{\theta}dt-dq_{\theta})$, where $\tau_d^{-1}$ is the detection bandwidth.     

Between triggers, the un-normalized conditioned state $|\overline{\psi}_{\rm REC}\rangle$ satisfies now the following Stochastic Schr\"odinger Equation (SSE)~\cite{Carmichael1993QTIII, Reiner2001, CarmichaelBook2}:
\begin{equation}\label{eq:SSEmain}
d|\overline{\psi}_{\rm REC}\rangle=\left(-\kappa a^{\dagger}a\,dt +\sqrt{2\kappa (1-r)}\, a\,e^{-i\theta} d\xi\right)|\overline{\psi}_{\rm REC}\rangle,
\end{equation}
where
\begin{equation}\label{eq:dximain}
\begin{aligned}
d\xi &\equiv e^{\kappa t}(G|\mathcal{E}_{\rm lo}|)^{-1}\,dq_{\theta}\\
&=\sqrt{2\kappa (1-r)}[(e^{i\theta}\langle a^{\dagger}  \rangle_{\rm REC} + e^{-i\theta} \langle a \rangle_{\rm REC} ) dt] + dW.
\end{aligned}
\end{equation}
Here, $G$ is a generalized circuit gain and $dW$ is a Gaussian-distributed random variable with zero mean and variance $dt$ (a Wiener increment). The two averages in Eq.~\eqref{eq:dximain} are to be calculated with respect to the normalized conditioned state
\begin{equation}
|\psi_{\rm REC}(t)\rangle=\frac{|\overline{\psi}_{\rm REC}(t)\rangle}{\sqrt{\langle \overline{\psi}_{\rm REC}(t)|\overline{\psi}_{\rm REC}(t)\rangle}}.
\end{equation}
Since the reservoir is in the vacuum state, the conditioned average field measured at the APD is proportional to $\sqrt{2\kappa r}\,\langle a(t)\rangle_{\rm REC}$, while the sample making is triggered with a probability equal to $2\kappa r \langle \psi_{\rm REC}(t)| a^{\dagger}a|\psi_{\rm REC}(t)\rangle\,dt$. The cumulative (or integrated) charge deposited in the detector is defined as the real number
\begin{equation}\label{eq:Qmain}
Q_{\theta} \equiv \sqrt{2\kappa} (G|\mathcal{E}_{\rm lo}|)^{-1}\int_{0}^{t}dq_{\theta}=\sqrt{2\kappa}\int_{0}^{t}e^{-\kappa t^{\prime}}d\xi^{\prime},
\end{equation}
an explicitly stochastic quantity. Carmichael has demonstrated~\cite{CarmichaelSG1994, CarmichaelBook2} that the probability distribution $P(Q_{\theta})$ over an ensemble of realizations in the limit $t\to \infty$ and for $r=0$ measures a marginal of the Wigner distribution representing in phase space the state of the cavity field immediately before the period of free decay. In this case $P(Q_{\theta})$ arises as steady-state solution to a Fokker--Planck equation, whose form depends on the quadrature amplitude selected by the LO phase $\theta$ (see App.~\ref{App:A}). 

Realizations of $I_{\theta}(t)$, the set of APD trigger times $\{t_j\}$ and the conditioned state $|\overline{\psi}_{\rm REC}(t)\rangle$ obey a set of stochastic differential equations than can be simulated on a computer via a Monte Carlo algorithm in the general case. In the specific example of a damped macroscopic superposition, we find that the stochastic dynamics can be formulated in a semi-analytical fashion without introducing a Hilbert space for the cavity mode (see App.~\ref{App:B}).

\subsection{The transient intensity-field correlation function}

By sampling an ongoing realization of the quadrature amplitude $A_{\theta}(t)$ for several ``start'' times $t_j$, $j=1,\ldots,N_s$, we can calculate an intensity-field correlation function~\cite{Carmichael2000, Foster2000,Reiner2001,Denisov2002,Wiseman2002,CarmichaelFosterChapter} as the following {\it transient} conditioned average~\cite{Carmichael1993QTIII, CarmichaelBook2},
\begin{equation}\label{eq:hdef}
h_{\theta}(t=0;\tau)\equiv \frac{1}{N_s}2\sqrt{2\kappa (1-r)}\,\sum_{j=1}^{N_s}\langle A_{\theta}(t_j+\tau)\rangle_{\rm REC},
\end{equation}
where $\langle A_{\theta}(t_j+\tau)\rangle_{\rm REC}=\langle \psi_{\rm REC}(t_j+\tau)|A_{\theta}|\psi_{\rm REC}(t_j+\tau)\rangle$ is a conditioned average along a single realization. The sum over $j$ in Eq.~\eqref{eq:hdef} is evaluated as an average over past and future measurement samples, before and after $t_j$~\cite{Carmichael1997}. The definition differs from the average photocurrent defined in~\cite{Carmichael2000,Foster2000,Reiner2001} because the number of samples (starts) available along a single trajectory is determined by $N_s=A^2$, which is not sufficient to reduce the shot noise appreciably when $A^2$ is of the order of the detection bandwidth (in units of $\kappa$). Furthermore, $h_{\theta}(0; \tau)$ is to be evaluated with reference to the initial coherent superposition state at $t=0$, instead of the steady state. For large-amplitude cats, we define $h_{\theta}(0;\tau)=(1/N_s)\sum_{j=1}^{N_s}I_{\theta}(t_j+\tau)$, since there are enough samples to recover the signal out of the shot noise. We expect on average $N_s=rA^2$ APD trigger ``clicks'' along any trajectory. Note that the ME~\eqref{eq:ME} predicts $\langle A_{\theta}(t) \rangle \sim 2 \sin\phi_0 \sin\theta \exp(-2A^2)$ as an ensemble average over different records with initial state~\eqref{eq:incat}, which entails a vanishingly small average $\langle h_{\theta}(0;\tau) \rangle$ for large initial photon numbers.

\section{Charge distribution and the phase $\phi_0$ of the initial superposition}
\label{sec:chdist}

Direct detection we briefly met in Sec.~\ref{sec:MEDD} stands out as a special case in the limit $r=1$, destined to assess the corpuscular nature of the scattered light. We will now operate in the opposite limit by setting $r=0$ to produce an uninterrupted continuous photocurrent at the BHD, thus focusing on the wave aspect of radiation. In Fig.~\ref{fig:FIG1}, we explore the effect of a varying phase between the two components in the initial superposition~\eqref{eq:incat} to the cumulative charge released by the detector in the course of the {\it entire} evolution to the vacuum, for $\theta=\pi/2$. The Wigner function of the initial cavity state is~\cite{Buzek1992, CarmichaelSG1994, Wodkiewicz2000, SchleichCh3, SchleichCh11, HarocheBook}
\begin{equation}\label{eq:W0}
\begin{aligned}
&W(x,y;t=0)=(2\pi e^{-A^2}\cosh A^2)^{-1} e^{-2y^2}\\
&\times [e^{-2(x-A)^2} + e^{-2(x+A)^2}+2e^{-2x^2}\cos(\phi_0 + 4Ay)],
\end{aligned}
\end{equation}
where the last term (cosine) indicates quantum interference~\cite{Zurek2001}. A contour plot of $W(x,y;t=0)$ for $\phi_0=0$ is given in inset (i). The Wigner function of the cavity state maintains the form of Eq.~\eqref{eq:W0} with $A\to A(t)\equiv A e^{-\kappa t}$ throughout the evolution dictated by ME~\eqref{eq:ME} but, crucially, the cosine term is scaled by the factor $\exp[-2A^2(1-e^{-2\kappa t})]$: the greater the initial distance of the two components the faster the off-diagonal elements of $\rho(t)$ are dephased~\cite{Walls1985, Phoenix1990,Zurek2003,Zurek2003B}. The disparity between the weights of the Gaussian peaks and the interference fringes brings in the short timescale $(2\kappa A^2)^{-1}$ as the relevant decoherence time we met in Sec.~\ref{sec:MEDD}. Simultaneously, the Wigner function maintains its symmetry with respect to the $x$- and $y$-axis in all stages of the decay, and so do its corresponding marginals. 

Setting $\theta=\pi/2$, the marginal distribution---obtained by integrating $W_{\rm REC}(x, y;t=0)$ along the $x$-axis---reads
\begin{equation}\label{eq:marginal}
P(y;t=0)=(\sqrt{2\pi}\cosh A^2)^{-1}e^{-2y^2+A^2} [1+\cos(\phi_0+4Ay)].
\end{equation}
The charge distribution $P(Q_{\pi/2})$ depicted in insets (ii-iv) is obtained {\it after} the light has left the cavity in the decay of an ensemble of scattering records to the vacuum. In Sec.~\ref{sec:condintf}, we will put this decay into a dynamical context with reference to the timescale $\kappa t_m=(1/2)\ln(2A^2)$ required for phase ``localization'' across individual realizations of $Q_{\pi/2}$, and for interference fringes to appear in their ensemble average. The distribution $P(Q_{\pi/2})$ and the marginal~\eqref{eq:marginal} are related by a simple scale factor with $Q_{\pi/2}=2y$. For all values of $\phi_0$ different to $0$ and $\pi$, $P(Q_{\pi/2})$ is asymmetric with respect to the $Q_{\pi/2}=0$-axis, yet the average deposited cumulative charge remains zero, as expected from the vanishing integral $\int_{-\infty}^{\infty} y P(y;t=0)dy$. The probability of depositing $Q_{\pi/2}$ in the vicinity of zero scales as $2\cos^2(\phi_0/2)$---a direct observational consequence of the initial phase. Therefore, mode-matched balanced homodyne detection performs a phase-sensitive tomogram~\cite{Smithey1993, Lutterbach1997, Welsch1999, SchleichCh4, Vogel2006, HarocheBook} of the initial cavity state. In contrast, as we have previously remarked, the interference term in the Wigner function corresponding to $\rho(t)$ evolving from the initial state~\eqref{eq:incat} under the action of the ME, disappears fast after the lapse of the decoherence time $(2\kappa A^2)^{-1}$ and, together with it, any remain of the initial phase difference between the two components in the coherent-state superposition.

\begin{figure*}
\includegraphics[width=\textwidth]{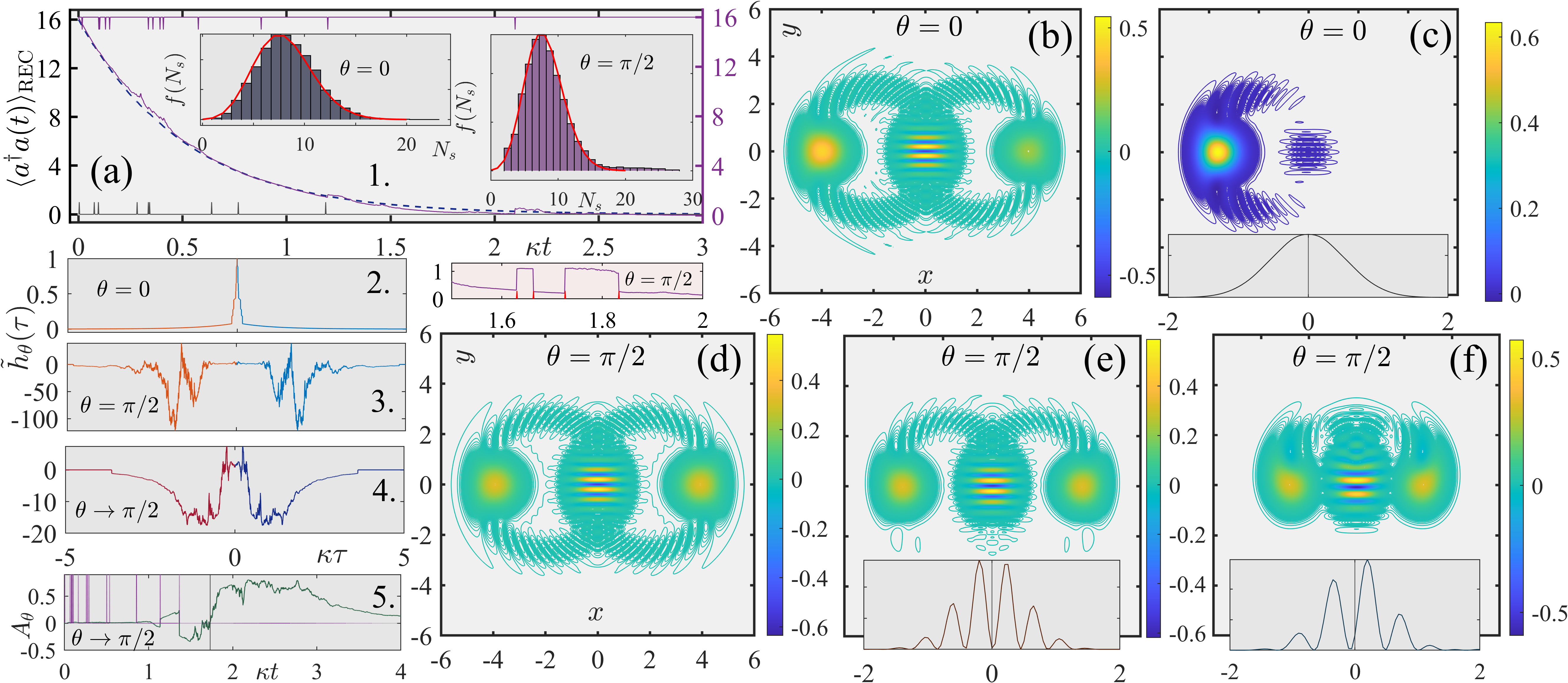}
\caption{Conditioned Monte Carlo averages and state representations in single realizations. {\bf (a)} {\bf 1}: Conditioned intracavity photon number $\langle a^{\dagger}a(t) \rangle_{\rm REC}$ against three cavity lifetimes, for $\theta=0$ (dashed line) and $\pi/2$ (solid line). The strokes underneath ($\theta=0$) and above ($\theta=\pi/2$) the main plots indicate photon emissions triggering the homodyne current generation at the BHD. The two insets depict relative frequency [$f(N_s)$] histograms of APD photon ``click'' resets, collected for $\theta=0$ (left) and $\theta=\pi/2$ (right), over 4,000 realizations. The curves in red depict the Poisson probability density $p(N_S)=\lambda^{N_s}e^{-\lambda}/N_s!$ for $\lambda=rA^2=8$. The small pink-shaded rectangular frame underneath shows a close-up of a different sample trajectory $\langle a^{\dagger}a(t) \rangle_{\rm REC}$ generated with $\theta=\pi/2$, in the course of half a cavity lifetime. The four red strokes underneath denote APD ``clicks''. {\bf (a)} {\bf 2 -- 4}: Individual realizations of the intensity-field correlation function over its zero-delay value, $\tilde{h}_{\theta}(\tau)\equiv h_{\theta}(0; \tau)/h_{\theta}(0; 0)$, for $\theta=0$ in (2), $\pi/2$ in (3) and $(\pi/2-\theta)=0.0078$ in (4). Frame (a5) depicts the conditioned average of the quadrature amplitude $A_{\theta}\equiv \langle A_{\theta}(t) \rangle_{\rm REC}$ corresponding to the trajectory of (a4). The dotted strokes in (a5) mark photon triggers, while the long vertical line marks the time $\kappa t_m=(1/2)\ln(2A^2)\approx 1.733$. {\bf (b)} Contour plot of the conditioned Wigner function $W_{\rm REC}(x+iy;t_1)$, at the time $\kappa t_1=0.004$ of the first photon trigger, along the trajectory of (a) generated for $\theta=0$; {\bf (c)} Similar to (b), with $W_{\rm REC}(x+iy;t_2)$ at the time $\kappa t_2=0.096$ of the third photon trigger. The lower inset depicts a numerical approximation of the conditioned marginal distribution $P_{\rm REC}(y;t_2)=\int_{-\infty}^{\infty} W_{\rm REC}(x+iy;t_2)\,dx$. {\bf (d)} Contour plot of the conditioned Wigner function $W_{\rm REC}(x+iy;t_1^{\prime})$, at the time $\kappa t_1^{\prime}=0.018$ of the first photon trigger, along the trajectory of (a) generated for $\theta=\pi/2$; {\bf (e)} Similar to (d), with $W_{\rm REC}(x+iy;t_2^{\prime})$, at the time $\kappa t_2^{\prime}=0.098$ of the second photon trigger; {\bf (f)} Similar to (d), with $W_{\rm REC}(x+iy;t_3^{\prime})$, at the time $\kappa t_3^{\prime}=0.392$ of the eighth photon trigger along the trajectory generated for $\theta=\pi/2$. The lower insets in (e, f) depict numerical approximations of the conditioned marginal distributions $P_{\rm REC}(y;t_k^{\prime})=\int_{-\infty}^{\infty} W_{\rm REC}(x+iy;t_k^{\prime})\,dx$ for $k=2, 3$, respectively. In all realizations, the initial state~\eqref{eq:incat} has $A=4$ and $\phi_0=0$, while the correlator operates with $r=0.5$. The time step size is $\kappa \Delta t=0.002$, and the Fock-state basis is truncated at the $30$-photon level; see App.~\ref{App:B2}.}
\label{fig:FIG2}
\end{figure*}

\section{Conditioned intensity-field correlations for ``position'' and ``momentum'' measurements}
\label{sec:condintf}

Let us now meet further evidence on how individual Monte Carlo realizations~\cite{Carmichael1993QTI} under the action of the wave-particle correlator (see also App.~\ref{App:B2}) subvert the picture offered by the ME~\eqref{eq:ME} and the Wigner function of the cavity state~\eqref{eq:W0} formulated as a statistical mixture over an ensemble of pure states. Figure~\ref{fig:FIG2} depicts results obtained when the correlator operates with $r=0.5$. The pair of sample trajectories in frame (a1) show a decaying conditioned intracavity photon number $\langle a^{\dagger}a(t) \rangle_{\rm REC}$ for the same input state and two different settings of the LO phase, $\theta=0$ and $\theta=\pi/2$. No large differences are noted between the two records, perhaps apart from some collapses leading to higher conditioned photon emission probability deviating from the exponential decay in three instances. The trend of such instability is visible in the histogram of total photon counts $N_s$ recorded along a given trajectory, computed over thousands of realizations. The distribution displays a longer tail when $\theta$ is set to $\pi/2$ as opposed to $\theta=0$, and marks a clear departure from a Poisson distribution with mean photon number $rA^2$. Additional APD trigger ``clicks'' are registered in individual realizations generated with $\theta=\pi/2$, often past the average photon lifetime, abruptly increasing the conditioned photon emission rate $2\kappa r \langle \psi_{\rm REC}(t)| a^{\dagger}a|\psi_{\rm REC}(t)\rangle$ and marking a departure from its expected exponential decay pictured in frame (a1).

The time-symmetric intensity-field correlations plotted in frames (a2--a4) are the first quantities we meet that point to a clear operational disparity between two of the complementary unravelings. Since the ensemble-averaged field ${\rm tr(\rho(t) A_{\theta})}$ [obtained from the solution of the ME~\eqref{eq:ME}] is zero for $\theta=0$, we expect conditioned correlation functions with positive peaks to cancel those with negative over an ensemble of realizations. Sharply decaying intensity-field correlations for $\theta=0$ gradually transition to highly oscillatory functions with alternating sign and notable deviations from their zero-delay values as $\theta\to\pi/2$. For the latter setting, there is also a notable difference between the intensity-field functions obtained for different realizations, testifying to another manifestation [in addition to the one of Fig.~\ref{fig:FIG2}(a1)] of the instability reported in~\cite{Carmichael1999}. With every photon trigger, a phase change of $\pi$ is generated between the two components of the cat state. However, not all triggers lead to a phase change in the conditioned field amplitude. The last trigger is the one to direct the field quadrature to one of the periodic wells of the modulated potential governing the evolution of $Q_{\theta}$~\cite{Carmichael1999} through the drift term of a Fokker--Planck equation (for the distribution of the integrated charge $Q_{\theta}$) and its equivalent SSE (for individual realizations).

Pronounced deviations are routinely observed past the time $\kappa t_m=(1/2)\ln(2A^2)\approx 1.733$, which is the time required for the potential to develop a deep periodic modulation (see App.~\ref{App:A2}). In Fig.~\ref{fig:FIG2}(a3), for example, a large oscillation of the field amplitude occurs within a particular potential well, dictating the frequency at which the conditioned field amplitude oscillates. The period and phase of this oscillation depend on the closeness of $\theta$ to $\pi/2$, as well as on the past phase diffusion between the two components, which determines whether a sign change occurs or not after a photon is recorded at the APD. In Figs.~\ref{fig:FIG2}(a4--a5), we meet a sample intensity-field correlation and the corresponding realization of field amplitude, respectively, calculated for $14$ resets (well in excess of $rA^2=8$) when $\theta \to \pi/2$. At $\kappa t_m$, a large excursion of the field amplitude is initiated after the last photon trigger in the series. The last reset following an APD ``click'' resolves the accumulated diffusion and is responsible for a sign change in the field amplitude. We note here the sensitivity of the dynamics to the value of $\theta$; a small deviation from $\theta=\pi/2$ produces a steady-state potential of rapidly decaying well heights with increasing $Q_{\theta}$ (see also App.~\ref{App:A2}), working against the localization into a particular well in the presence of shot noise. The intense fluctuations in the field amplitude visible in Figs.~\ref{fig:FIG2}(a4--a5) testify to a weaker localization.

We now move to the phase-space representation of the conditioned cavity states. Wigner functions of the cavity state conditioned on photon triggers, calculated for the pure system state  $\rho_{\rm REC}(t)=|\psi_{\rm REC}(t)\rangle \langle \psi_{\rm REC}(t)|$ (through the algorithm detailed in App.~\ref{App:B2}), are more ostensibly at odds with the ensemble-averaged profile described by Eq.~\eqref{eq:W0}. We first measure the ``position'' of the oscillator ($\theta=0$) prepared in the superposition state~\eqref{eq:incat}. The contour plot depicted in Fig.~\ref{fig:FIG2}(b) corresponds to the state collapse following a ``click'' occurring at a time which is an order of magnitude shorter than the decoherence time $(2\kappa A^2)^{-1}$ predicted by the ME~\eqref{eq:ME}. The interference fringes are in place, although there is a visible asymmetry in the amplitude of the side Gaussian peaks. With the lapse of about three decoherence times, past the value $(2\kappa r A^2)^{-1}$, the right peak has completely disappeared [Fig.~\ref{fig:FIG2}(c)], leaving behind only insignificant trails of quantum interference. From that point onwards, the evolution essentially concerns the decay of a single coherent state with a peak centred at $-A e^{-\kappa t}$ in phase space. For other trajectories generated with $\theta=0$, the single peak in the phase-space profile is instead centred at $A e^{-\kappa t}$. Therefore, the conditioned states produced for an unraveling with $\theta=0$ challenge the decoherence picture offered by the ME~\eqref{eq:ME} through a fast-developing unbalance between the two state components. Past the decoherence time, the statistics of the photon resets---the vast majority of the recorded APD ``clicks''---align with a decaying coherent state. An unbalance of similar kind is also met in the direct-photodetection unraveling of the ME~\eqref{eq:ME}, where the times of photoelectron counts bring into play a dynamical competition for an initial superposition state of different amplitudes~\citep{CarmichaelSG1994, Carmichael2013Ch4}.

Measuring the ``momentum'' of the harmonic oscillator ($\theta=\pi/2$) restores the interference in the conditioned Wigner distributions at all times [Fig.~\ref{fig:FIG2}(d--f)], present even when the damped cavity mode contains half of its initial photons [Fig.~\ref{fig:FIG2}(f)]. Phase diffusion over an ensemble of such states is responsible for a nearly uniform quantum phase distribution past the very short decoherence time $(2\kappa A^2)^{-1}$.  Moreover, the previous asymmetry with respect to the $y$-axis, is now developing with respect to the $x$-axis and distorts the interference, a trait also reflected in the conditioned marginals $P_{\rm REC}(y)$ obtained by integrating the Wigner function of the cavity state conditioned on an APD ``click''. Photon triggers interrupt the otherwise continuous phase diffusion by injecting a $\pi$-phase difference between the two components of the cat state, in a similar fashion to the $(-1)^n$ factor in Eq.~\eqref{eq:psiDP}, as the cat state leaves the cavity to partly exist in the output field. The interference fringes in Figs.~\ref{fig:FIG2}(d--f) resolve the phase change. All photon emission times corresponding to the conditioned Wigner functions in Figs.~\ref{fig:FIG2}(d--f) are well below the time $\kappa t_m$. Upon approaching that time, we expect phase localization of the cumulative charge trajectory to a particular potential well. Since for the vast majority of sample realizations we have $\langle a^{\dagger}a(t \sim t_m)\rangle_{\rm REC}<1$ for every $A^2 \gg 1$, this timescale also marks the eventual degradation of fringe visibility in the conditioned cavity Wigner function while the two coherent-state peaks decay to the phase-space origin. Nevertheless, rare deviations from this trend do occur.

To place these deviations in context, it is first instructive to compare the markedly different dynamical evolution of the conditioned cavity state when measuring ``position'' and ``momentum'', to the quantum state diffusion of a qubit, initialized in a superposition state, with selective quadrature measurement of a dispersively coupled cavity field. In the circuit QED experiment reported in~\cite{Qubit2013}, trajectories were confined either to the equator or a meridian of the Bloch sphere, depending on whether the amplified quadrature of the cavity field conveyed information on the qubit state, or whether it encoded intracavity photon number fluctuations shifting the phase of the qubit state. In the latter case, persistent oscillations of the polarization were recorded against the induced phase shift, with the inversion remaining uncorrelated with the measured signal. The confinement to the equator of the Bloch sphere certainly reflects the probed coherence of the initial pure-state superposition, as does the persistence of interference fringes when setting $\theta=\pi/2$ in the wave-particle correlator; $|\psi_{\rm REC}(t)\rangle$ remains a superposition of coherent states at all times.  The quantum state diffusion in the contextual decay of a cat state when setting $\theta=\pi/2$, however, is governed by a highly nontrivial equation, nonlinearly depending on both time $t$ and amplitude $A$ (intracavity excitation) as we have already discussed. Instead of the random walk in the angular diffusion of the probability density matrix observed in the circuit QED experiment of~\cite{Qubit2016} measuring non-commuting observables, we are now dealing a diffusion process~\cite{Carmichael1999} solving an SSE with an initially flat potential eventually to be succeeded by a deep periodic modulation at a sub-photon cavity occupation, responsible for phase localization at $t_m$. Only after this succession has seen operational consequences can we speak of a steady state at which fringes are formed in the measured signal---the cumulative charge.

Perhaps the most illustrative example of the `tension' between particles and waves when measuring ``momentum'' is the one noted in parallel with the aforementioned degradation of the fringe visibility in the cavity Wigner function close to $t_m$: a relative $\pi$ phase shift between the two components of the coherent-state superposition, induced by a photon emission (particle), works against the phase localization due to the monitoring of one quadrature phase amplitude (wave). Such instability is correlated with the large excursions of the field amplitude past $t_m$ as well as with departures from the Poisson distribution of the APD photoelectron ``clicks'' [see Fig.~\ref{fig:FIG2}]. Several trajectories of the conditional transmitted photon flux, $2\kappa \langle a^{\dagger}a(t) \rangle_{\rm  REC}$, visibly deviate from a monotonic exponential decay in the vicinity of $t_m$. In the absence of registered APD ``clicks'', it often exhibits a local maximum or a plateau about $t_m$ with a periodic modulation. The latter reflects the charge oscillation about a certain potential well (phase localization) whose depth increases with time.

For instance, in recognition of two closely-spaced photoelectron ``clicks'' in the small coloured frame underneath Fig.~\ref{fig:FIG2}(a1), the conditioned emission probability increases to operationally substantiate photon bunching~\cite{Carmichael1993QTII}---ensuring the second photon from each pair will be emitted a short time after the first---during an interval $\Delta t \ll (2\kappa)^{-1}$ along the trajectory, centered at $t_m$. In this particular realization, 18 APD ``clicks'' are registered in total (only the last four are shown), in what reinforces the deviation from the Poisson distribution of mean $\lambda=rA^2=8$ as a long tail of the corresponding right histogram in Fig.~\ref{fig:FIG2}(a1). Furthermore, we note that this type of photon bunching about $t_m$ disappears for the trajectories generated with $\theta=0$, or equally for those unraveling the ME of a decaying cavity mode initialized in a single coherent state. Notwithstanding these intricacies, the {\it ensemble average} of the conditioned cavity states $|\psi_{\rm REC}(t)\rangle \langle \psi_{\rm REC}(t)|$ over all realizations, produced by the wave-particle correlator operating at $\theta=\pi/2$, formally obeys one and the same deterministic ME~\eqref{eq:ME} (as it does for any value of $\theta$) rather than following a stochastic evolution which depends on the selected quadrature~\cite{Qubit2016}.

Having explored key differences in the correlations between the APD trigger ``clicks'' and the cavity state depending on the monitored quadrature phase, both in the time domain and in the phase-space representation, we may ask the question how is the electromagnetic field amplitude fed back to the photon emission events that condition it. Instead of laying down the entire photon emission sequence as we did in Fig.~\ref{fig:FIG2}(a1), in Sec.~\ref{sec:photonemdiff} we will restrict our attention to the first pair of the APD trigger ``clicks'' positioned against their waiting-time distribution, another key quantity attributed to the corpuscular nature of the scattered light.             

\section{Photon emissions as diffusion markers}
\label{sec:photonemdiff}

Owing to the consistent coupling between a classical and a quantum stochastic process accomplished by quantum trajectory theory~\cite{Carmichael1993QTI, Carmichael1999,Gambetta2008, Qubit2013,Qubit2018, Minev2019}, we can derive semi-analytical formulas connecting the charge production in the BHD and the trigger rate (see App.~\ref{App:B1} for further details). As we have already seen, central to the evolution between the triggers is a diffusion process either in the relative amplitude ($\theta=0$) or phase ($\theta=\pi/2$) [or a combination of the two for any other value of $\theta$] between the two components of the cat state~\eqref{eq:incat}, encapsulated in the null-measurement record, i.e. with no registered APD trigger ``click'':
\begin{equation}\label{eq:psiNULL}
\begin{aligned}
|\overline{\psi}&_{\rm REC,\, NULL}(t)\rangle=\exp[Q_{\theta}(t)A\sqrt{1-r}\,e^{-i\theta}]\,|A e^{-\kappa t} \rangle\\
& + \exp(i\phi_0)\exp{[-Q_{\theta}(t)A\sqrt{1-r}\,e^{-i\theta}]}\,|-A e^{-\kappa t} \rangle.
\end{aligned}
\end{equation}
For $r=0$ we recover the {\it ansatz} in Eqs. (22-23) of~\cite{CarmichaelSG1994} for balanced mode-matched homodyne detection, while at $t=0$, there is no net charge released from the detector ($Q_{\theta}=0$), whence we arrive at the un-normalized version of the initial state~\eqref{eq:incat}. For $r\to 1$, diffusion does not participate in the conditioning and we reach Eq.~\eqref{eq:psiDP} as direct detection is performed in the one arm left of the correlator~\footnote{The common pre-factor omitted between the two components must be reinstated to interpret \unexpanded{$\langle \overline{\psi}_{\rm REC,\, NULL}(t)|\overline{\psi}_{\rm REC,\, NULL}(t)\rangle$} as a record probability density.}. 

The central difference between the coefficient ratio of the decaying coherent states in Eqs.~\eqref{eq:psiDP} and~\eqref{eq:psiNULL} lies on the presence of a complex-argument exponential involving the cumulative charge in the place of unity in Eq.~\eqref{eq:psiDP} for $n=0$. This instance highlights a diffusion process with an inherent stochasticity instead of the deterministic evolution characterizing a null-measurement record in direct photodetection. For $\theta=0$ the two exponents remain real throughout the evolution. During a single realization, $Q_{\theta}$ is rapidly directed to either side of a $\Lambda$-shaped potential, which means that one component of the superposition wins over the other and creates the imbalance we met in Sec.~\ref{sec:condintf} (see App.~\ref{App:A2}). On the other hand, for $\theta=\pi/2$, the exponents in~\eqref{eq:psiNULL} are purely imaginary, and the two components acquire a phase difference conditioned on the charge production at the BHD; their relative weight equals unity and is not affected by the phase diffusion.  
\begin{figure}
\includegraphics[width=0.45\textwidth]{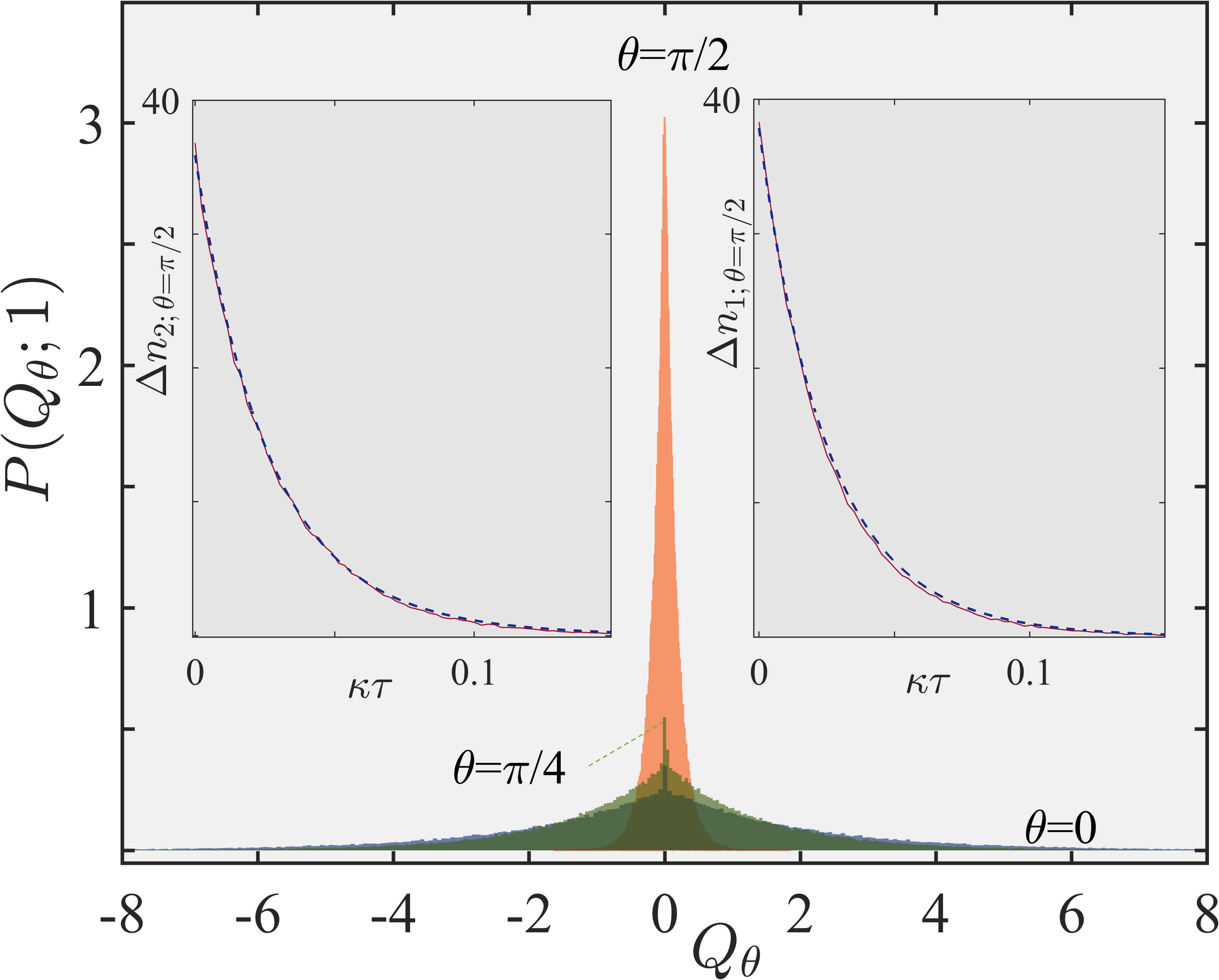}
\caption{Normalized distributions (histograms) of the net charge $Q_{\theta}$ deposited in the BHD at the time $t_1$ of the first photon trigger ``click'', for three different settings of the LO phase, $\theta=0$, $\theta=\pi/4$ and $\pi/2$, indicated accordingly. The two insets either side of the histograms depict distributions of waiting times for the first (right) and second (left) photon emissions. $\Delta n_1$ ($\Delta n_2$) is the number of times photon 1 (photon 2) is emitted in the time interval $[\tau,\tau+\Delta\tau)$~\cite{CarmichaelKim2000}. The continuous lines plot an ensemble average of 150,000 trajectories with bin size $\kappa \Delta \tau=0.0025$, while the dashed lines correspond to Eq.~\eqref{eq:wapprox} with $A$ (right) and $A\to A e^{-\kappa \tau_{\rm av}}$ (left). The initial state has: $A=20$, $\phi_0=\pi$, while the correlator operates with $r=0.05$.}
\label{fig:FIG3}
\end{figure} 

We focus on higher-amplitude cat states, such as those generated via conditional qubit-photon logic in circuit QED~\cite{Vlastakis2013, Girvin2019}, to produce a large number of closely-spaced photon triggers. We operate the correlator with $r\ll 1$ to approach a pure balanced homodyne detection, yet satisfying $rA^2 \gg 1$. The waiting-time distribution~\cite{Carmichael1989, CarmichaelKim2000, Brandes2008} for the first photon trigger, a characteristic particle-type attribute, is approximated by the analytical expression
\begin{equation}\label{eq:wapprox}
w_1(\tau)\approx 2\kappa \frac{\exp[-rA^2(1-e^{-2\kappa \tau})]}{\int_0^1 \exp[-rA^2(1-e^{-u})]du},
\end{equation}
in line with a no-``click'' measurement record obtained for a decaying coherent state with initial amplitude $\sqrt{rA^2}$~\citep{Carmichael2013Ch4}. The average time waited until the first photon emission is $\tau_{\rm av}= (2\kappa rA^2)^{-1}$. The two insets of Fig.~\ref{fig:FIG3} show that the emission times of the first photon and the second, conditioned on the first reset, follow Eq.~\eqref{eq:wapprox}---the latter with the replacement $A\to A e^{-\kappa \tau_{\rm av}}$---when the ``momentum'' of the oscillator is measured. We find that the same trend is followed when measuring ``position''; the obtained waiting-time distributions overlap with those shown in the figure. 

The cumulative charge diffusion conditioned on a photon detection and registered at the BHD, however, is very different in these two settings, as we can observe in the main plots of Fig.~\ref{fig:FIG3}. Within the average photon emission waiting time, the distribution of $Q_{\theta}$ diffuses at a similar rate when $\theta=0$ and $\theta=\pi/4$, while a pronounced and disproportionate difference is noted for $\theta=\pi/2$. Considering that waves (quadrature amplitudes) condition the emission of particles (photons), we deduce from Eqs.~\eqref{eq:dximain} and~\eqref{eq:Qmain} that the conditioned electromagnetic field amplitude will vary at a slower rate when measuring ``momentum''. For the latter setting, the charge distribution is found to remain virtually unchanged when conditioned on the second photon trigger as well. Finally, we remark that $\kappa \tau_{\rm av} \ll (1/2)\ln(2A^2)$, whence no fringes are created in the charge distributions of Fig.~\ref{fig:FIG3}, like those we discussed in Sec.~\ref{sec:chdist}. 

\section{Concluding remarks} 

In summary, we have shown that: {\bf (i)} the steady-state solution of a Fokker-Planck equation reflects the statistical behaviour of the cumulative charge produced by mode-matched balanced homodyne detection. The charge distribution drastically changes with the initial phase between the two components of a macroscopic quantum state superposition (Sec.~\ref{sec:chdist}); {\bf (ii)} even for moderate intracavity excitation ($A\gtrsim 1$), the stability in the stochastic dynamics associated with measuring ``position'' of the damped cavity mode is accompanied by a rapidly developing asymmetry in the relative weight of the two state components while, when measuring ``momentum'', the associated instability manifests in large relative deviations of the intensity-field correlation function with a characteristic long timescale. While the cavity remains in a superposition of coherent states at all times, it is the above timescale that determines the appearance of interference fringes in the ensemble-average distribution of cumulative charge records (Sec.~\ref{sec:condintf}). The underlying phase localization is linked to an instability occasionally resolved as a rare fluctuation with a $\pi$ relative phase change arising when a photoelectron ``click'' is registered at the APD; {\bf (iii)} photon emissions from the cavity can be used to mark the extent by which the cumulative charge has diffused by amplitude and/or phase, which is indicative of the electromagnetic field variation inside the cavity (Sec.~\ref{sec:photonemdiff}). 

Our perspective has thus moved from {\it a posteriori} inferring a set initial phase difference from an ensemble of realizations to measuring a dynamical and complementary amplitude and phase diffusion along {\it single realizations}, in an unraveling method which takes both the particle and wave aspects of the scattered light into account. Notable differences in wave-particle correlations across complementary unravelings can be detected even for low-amplitude optical cat states subject to the current experimental limitations~\cite{Nielsen2006, Ourjoumtsev2006,Ourjoumtsev2007, Takahashi2008, Huang2015, Ulanov2016, Sychev2017,Hacker2019,Takase2021,Li2024}. The `tension' between particles and waves illustrated in~\cite{Carmichael2001} is revealed via two distinct timescales. In the one extreme, when the local oscillator is tuned to measure ``position'', a strong unbalance between the two parts develops from the very start. By the lapse of the decoherence time required to turn the initial pure state into a statistical mixture in the ensemble average governed by the ME, the interference pattern effectively disappears while the cavity contains a significant amount of excitation. Most photons are subsequently emitted in the presence of a single coherent state in the cavity. On the other end, when the local oscillator is tuned to measure ``momentum'', the interference fringes make their appearance late, after $\ln (2 A^2)$ photon decay times, when the cavity output pulse has reached the tail of the exponential decay. Occasional conditional emission probability ``spikes'' occur as rare fluctuations in that timescale---the ones responsible for the long tail in the right histogram of Fig.~\ref{fig:FIG2}(a1). 

The aforementioned timescale separation in the decay of macroscopic coherent-state superpositions ($A \gg 1$) leaves no room for the quantum interference to influence the waiting-time distribution of the overwhelming majority of the emitted photons resetting the balanced homodyne detection. Nonetheless, photon emission times serve as diffusion markers of the charge generated at the homodyne detector, and of the electromagnetic field amplitude in the cavity. Such markers are to be applied through conditioned cavity state tomograms~\cite{Smithey1993, Lutterbach1997, Nogues2000, SchleichCh4, Deleglise2008,Hofheinz2008,Hofheinz2009,Eichler2012,Haroche2013,Blais2021,Ahmed2021,Ahmed2021B,Wang2022,He2023}, and are embedded in a highly contextual intensity-field correlation function.

\appendix
\onecolumngrid 
\par\noindent\rule{\textwidth}{0.5pt}
\setcounter{equation}{0}

In the following appendices, we detail the amplitude and phase diffusion of the conditioned state under the complementary wave-particle correlator unravelings. We derive a Fokker--Planck equation and an associated potential for mode-matched homodyne detection, and point to the link between the steady-state distribution and the marginal distribution of the initial Wigner function. We derive the general form of trajectories for conditional homodyne detection and, finally, delineate the steps to produce them via the corresponding Monte Carlo numerical procedure.

\section{Stochastic Schr\"odinger equation, Fokker--Planck equation and the associated potential in homodyne detection}
\label{App:A}

\subsection{Stochastic Schr\"odinger equation and its transformation}
\label{App:A1}

We wish to determine the evolution of the cat state~\eqref{eq:incat},
\begin{equation}
|\psi_{\rm REC}(0)\rangle=\frac{|A\rangle + e^{i\phi_0}|-A\rangle}{\sqrt{2[1+\cos\phi_0\exp(-2A^2)]}},
\end{equation}
conditioned upon the operation of the wave-particle correlator. Between photon triggers, corresponding to the action of the super-operator $2\kappa r a (|\overline{\psi}_{\rm REC}\rangle \langle \overline{\psi}_{\rm REC}|) a^{\dagger}$, the un-normalized conditioned state $|\overline{\psi}_{\rm REC}\rangle$ satisfies the following Stochastic Schr\"odinger Equation (SSE)~\cite{Carmichael1993QTIII, Reiner2001, CarmichaelBook2}:
\begin{equation}\label{eq:SSEsup}
d|\overline{\psi}_{\rm REC}\rangle=\left(-\kappa a^{\dagger}a\,dt +\sqrt{2\kappa (1-r)}\, a\,e^{-i\theta} d\xi\right)|\overline{\psi}_{\rm REC}\rangle,
\end{equation}
where
\begin{equation}\label{eq:dxisup}
d\xi \equiv e^{\kappa t}(G|\mathcal{E}_{\rm lo}|)^{-1}\,dq_{\theta}=\sqrt{2\kappa (1-r)}[(e^{i\theta}\langle a^{\dagger}  \rangle_{\rm REC} + e^{-i\theta} \langle a \rangle_{\rm REC} ) dt] + dW.
\end{equation}
In the above, $G$ is the detecting circuit gain coefficient; $dW$ is a Wiener increment with zero mean and variance $dt$. These equations govern the photocurrent production after taking into account the detection bandwidth.

To proceed we set~\cite{CarmichaelBook2} 
\begin{equation}
|\overline{\psi}_{\rm REC}\rangle=e^{-\kappa a^{\dagger}a\,t}|\chi\rangle,
\end{equation} 
which transforms Eq.~\eqref{eq:SSEsup} to
\begin{equation}
\begin{aligned}
d|\chi\rangle&=\sqrt{2\kappa (1-r)}\,e^{\kappa a^{\dagger}a t}\, a\, e^{-i\theta} e^{-\kappa a^{\dagger}a t}\,d\xi|\chi\rangle\\
&=\sqrt{2\kappa(1-r)}\,e^{-\kappa t}\,a\,e^{-i\theta}\,d\xi\,|\chi\rangle\\
&=a\,\sqrt{1-r}\,dQ_{\theta}e^{-i\theta}|\chi\rangle,
\end{aligned}
\end{equation}
with solution $|\chi\rangle=e^{a\,\sqrt{1-r}\, e^{-i\theta}Q_{\theta}}|\chi(0)\rangle=e^{a\,\sqrt{1-r}\, e^{-i\theta}Q_{\theta}}|\psi_{\rm REC}(0)\rangle$. Substituting for $|\psi_{\rm REC}(0)\rangle$ the initial state~\eqref{eq:incat}, we find
\begin{equation}
|\overline{\psi}_{\rm REC}(t)\rangle=\frac{\exp[-A^2(1-e^{-2\kappa t})]}{\sqrt{2[1+\cos\phi_0\exp(-2A^2)]}}\left\{\exp[Q_{\theta}A\sqrt{1-r}\,e^{-i\theta}]\,|A e^{-\kappa t} \rangle + \exp[i\phi_0-Q_{\theta}A\sqrt{1-r}\,e^{-i\theta}]\,|-A e^{-\kappa t} \rangle \right\}.
\end{equation}
The common prefactor is omitted in Eq.~\eqref{eq:psiNULL}, while for $r=1$ we obtain the null-measurement record for direct photodetection, as with setting $n=0$ in Eq.~\eqref{eq:psiDP}~\cite{Carmichael2013Ch4}.

Knowing the form of the system wavefunction in conditional homodyne detection, we can then evaluate the conditioned expectation of the quadrature amplitude until the first photon trigger:
\begin{equation}
\begin{aligned}
\sqrt{1-r}\langle A_{\theta}(t)\rangle_{\rm REC}&=\frac{1}{2}\sqrt{1-r}\,\frac{\langle \psi_{\rm REC}(0)| e^{\sqrt{1-r}\,Q_{\theta}a^{\dagger}e^{i\theta}} e^{-\kappa a^{\dagger}a t} a^{\dagger} e^{i\theta} e^{-\kappa a^{\dagger}a t} e^{\sqrt{1-r}\,Q_{\theta}a e^{-i\theta}}|\psi_{\rm REC}(0)\rangle + {\rm c.c.}}{\langle \psi_{\rm REC}(0)| e^{\sqrt{1-r}\,Q_{\theta}a^{\dagger}e^{i\theta}} e^{-\kappa a^{\dagger}a t} e^{-\kappa a^{\dagger}a t} e^{\sqrt{1-r}Q_{\theta}a e^{-i\theta}}|\psi_{\rm REC}(0)\rangle}\\
&=\frac{1}{2} e^{-\kappa t}  \frac{\partial}{\partial Q_{\theta}}\ln\left[\langle \psi_{\rm REC}(0)| e^{\sqrt{1-r}\,Q_{\theta} a^{\dagger}e^{i\theta}} e^{-2\kappa a^{\dagger}a t} e^{\sqrt{1-r}\,Q_{\theta}a e^{-i\theta}}|\psi_{\rm REC}(0)\rangle \right].
\end{aligned}
\end{equation}
Substituting this conditioned expectation to Eqs.~\eqref{eq:SSEsup} and~\eqref{eq:dxisup}, yields
\begin{equation}\label{eq:dQHCond}
dQ_{\theta}=-\frac{\partial}{\partial Q_{\theta}} V(Q_{\theta},t) (2\kappa\,e^{-2\kappa t}\,dt) + \sqrt{2\kappa} e^{-\kappa t}\, dW,
\end{equation}
where, in anticipation of a ``drift'' term in a Fokker--Planck equation, we have introduced the time-dependent potential
\begin{equation}\label{eq:potential}
V(Q_{\theta},t)=-\ln\left[\langle \psi_{\rm REC}(0)| e^{\sqrt{1-r}\, Q_{\theta} a^{\dagger}e^{i\theta}} e^{-2\kappa a^{\dagger}a t} e^{\sqrt{1-r}\,Q_{\theta}a e^{-i\theta}}|\psi_{\rm REC}(0)\rangle \right],
\end{equation} 
explicitly depending on the initial state. With the change of variable~\cite{CarmichaelBook2} $\eta=1-e^{-2\kappa t}$, Eq.~\eqref{eq:dQHCond} is transformed to
\begin{equation}\label{eq:SDEtr}
dQ_{\theta}=-\frac{\partial}{\partial Q_{\theta}} V(Q_{\theta},\eta) d\eta + d\zeta,
\end{equation}
where $d\zeta$ is another Wiener increment with zero mean and variance $d\eta$. 

After the first photon click at time $t_1$, the initial wavefunction $|\psi_{\rm REC}(0)\rangle$ is updated at $t_1+dt$ to 
\begin{equation}
\frac{a|\overline{\psi}_{\rm REC}(t_1)\rangle}{\langle \overline{\psi}_{\rm REC}(t_1)|a^{\dagger}a|\overline{\psi}_{\rm REC}(t_1) \rangle},
\end{equation}
and the potential~\eqref{eq:potential} is modified accordingly.

\subsection{Balanced mode-matched homodyne detection ($r=0$): Fokker--Planck equation and individual realizations}
\label{App:A2}

For $r=0$, there are no photon ``click'' resets, and we recover the potential corresponding to mode-matched homodyne detection. The treatment is considerably simplified since Eq.~\eqref{eq:SDEtr} applies throughout the evolution. Distributions are then obtained after generating an ensemble of single realizations solving the stochastic Eq.~\eqref{eq:SDEtr} with the transformed potential
\begin{equation}\label{eq:potentialtr}
V(Q_{\theta},\eta)=-\ln\{\cosh(2Q_{\theta}A\cos\theta)\exp[A^2(1-2\eta\cos^2\theta)] + \cos(\phi_0+2 Q_{\theta}A \sin\theta)\exp[-A^2(1-2\eta\sin^2\theta)]\}.
\end{equation}
The steady-state limit is taken $\eta\to 1$ ($t\to \infty$), and results are plotted in Fig.~\ref{fig:FIG1}.

For $\theta=0$, the potential has a $\Lambda$ shape throughout the evolution, and each realization of $Q_{\theta}(\eta)$ solving~\eqref{eq:SDEtr} is directed to either a positive or a negative value. This type of symmetry breaking is revealed by the vanishing peak in the Wigner function of Figs. 2 (b, c). For $\theta=\pi/2$, the potential remains flat until the second time-dependent factor in Eq.~\eqref{eq:potentialtr} approaches the order of magnitude of the first, with $\exp(-2A^2 e^{-2\kappa t_m}) \sim 1/e$. On approaching that time $t\sim t_m=(2\kappa)^{-1}\ln(2A^2)$ in the evolution, the potential develops a deep periodic modulation. The well heights are very sensitive to variations of $\theta$ about $\pi/2$. The conditioned quadrature amplitude attains then an appreciable value with respect to ME ensemble average, depending on $\langle A e^{-\kappa t}|-A e^{-\kappa t} \rangle=\exp(-2A^2 e^{-2\kappa t})$. This is the charge accumulation time required for the interference pattern to appear in the distribution $P(Q_{\theta=\pi/2})$ of the cumulative charge deposited in the BHD. All realizations of the cumulative charge used for Fig.~\ref{fig:FIG1} have progressed well past that time. In contrast, the average time waited until the first photon trigger in Fig.~\ref{fig:FIG3} (with $r \ll 1$) is $\tau_w=(2\kappa r A^2)^{-1} \ll t_m$, {\it a priori} precluding the appearance of any interference fringes in the conditioned transient $P(Q_{\pi/2};t_1)$, where $t_1$ is the time of the first photon ``click'' at the APD.      

The stochastic differential equation (SDE)~\eqref{eq:SDEtr} with $r=0$ is equivalent to the following Fokker--Planck equation for the charge distribution $P(Q_{\theta},\eta)$~\cite{CarmichaelSG1994}:
\begin{equation}
\frac{\partial P(Q_{\theta},\eta)}{\partial \eta}=\left(-\frac{\partial}{\partial Q_{\theta}} \left(\frac{\partial V(Q_{\theta},\eta)}{\partial Q_{\theta}} \right)+\frac{1}{2}\frac{\partial^2}{\partial Q_{\theta}^2}\right)P(Q_{\theta},\eta).
\end{equation}
By direct substitution, we find the solution in the form:
\begin{equation}
\begin{aligned}
P(Q_{\theta},\eta)=&[2\sqrt{2\pi \eta}\,\cosh(A^2)]^{-1}e^{-Q_{\theta}^2/(2\eta)}\\
&\times\left\{\cosh(2Q_{\theta}A\cos\theta)\exp[A^2(1-2\eta\cos^2\theta)] + \cos(\phi_0+2 Q_{\theta}A \sin\theta)\exp[-A^2(1-2\eta\sin^2\theta)] \right\}.
\end{aligned}
\end{equation}
With the passage of time, the above expression converges at different rates (depending on the LO phase) to the marginal of the Wigner function of the {\it initial} cavity state. The marginals are obtained integrating over the phase-space co-ordinate transverse to the direction of the phasor representing the local oscillator. The distribution should be rescaled with $Q_{\theta}=2(\cos\theta \,x +\sin\theta\, y)$ in the steady state ($\eta \to 1$), to be {\it a posteriori} identified with the inferred initial marginal. For $\theta=\pi/2$, we obtain Eq.~\eqref{eq:marginal}, namely the marginal
\begin{equation}
P(y;t=0)=(\sqrt{2\pi}\cosh A^2)^{-1}e^{-2y^2+A^2} [1+\cos(\phi_0+4Ay)],
\end{equation}
plotted in Fig.~\ref{fig:FIG1} and superposed on the histograms of the long-time limit in the realizations solving~\eqref{eq:SDEtr} with $r=0$, for different values of $\phi_0$.

\section{Conditioned wavefunction and Monte Carlo algorithm}
\label{App:B}

\subsection{Conditioned evolution and photon emission rates}
\label{App:B1}

The wave-particle correlator unraveling consists of a continuous homodyne current generation reset by photon ``clicks'' recorded by the APD. Putting the pieces together and, owing to the linearity of the SSE~\eqref{eq:SSEsup}, for the conditioned wavefunction we obtain the superposition:
\begin{equation}\label{eq:psia1a2}
|\overline{\psi}_{\rm REC}(t) \rangle=|\overline{\psi}^{(A)}_{\rm REC}(t) \rangle + |\overline{\psi}^{(-A)}_{\rm REC}(t) \rangle,
\end{equation}
with each component of the initial state following a different evolution for a trajectory with $n$ photon resets at the times $t_1, t_2,\ldots,t_n$:
\begin{equation}\label{eq:psibeta}
\begin{aligned}
&|\overline{\psi}^{(\beta)}_{\rm REC}(t) \rangle=\exp\left[-\tfrac{1}{2}|\beta|^2 (e^{-2\kappa t_n}-e^{-2\kappa t})\right][\sqrt{2\kappa r}\,\beta (t_n)]e^{\beta(t_{n-1})Q_{\theta;\,n}\sqrt{1-r}\,e^{-i\theta}}\ldots\\
&\ldots e^{\beta(t_1)Q_{\theta;\,2}\sqrt{1-r}\,e^{-i\theta}}\exp\left[-\tfrac{1}{2}|\beta|^2 (e^{-2\kappa t_1}-e^{-2\kappa t_2})\right][\sqrt{2\kappa r}\,\beta(t_1)]e^{\beta(0)Q_{\theta;\,1}\sqrt{1-r}\,e^{-i\theta}}\,\exp\left[-\tfrac{1}{2}|\beta|^2 (1-e^{-2\kappa t_1})\right]\,|\beta (t)\rangle,
\end{aligned}
\end{equation}
where $\beta(t)=\beta e^{-\kappa t}$ and $\beta=A, -A$. The cumulative charges $Q_{\theta;\,1}, Q_{\theta;\,2}, \ldots Q_{\theta;\,n}$ are stochastic quantities produced after each reset and correspond to the intervals $(t_1-0), (t_2-t_1),\ldots,(t_{n}-t_{n-1})$, respectively.

For $r=1$, we revert to:
\begin{equation}
|\overline{\psi}^{(\beta)}_{\rm REC}(t) \rangle=(\sqrt{2\kappa}\,\beta\,e^{-\kappa  t_n})\ldots (\sqrt{2\kappa}\,\beta\,e^{-\kappa t_1})\,\exp\left[-\tfrac{1}{2}|\beta|^2 (1-e^{-2\kappa t})\right]\,|\beta (t)\rangle,
\end{equation}
the familiar formula for direct photodetection~\cite{Carmichael2013Ch4}. Now the factor $\exp\left[-\tfrac{1}{2}|\beta|^2 (1-e^{-2\kappa t})\right]$ captures the no-``click'' evolution.

We can now apply Eqs.~\eqref{eq:psia1a2} and~\eqref{eq:psibeta} to determine the emission probabilities of the first two photons recorded by the APD, used for Fig.~\ref{fig:FIG3}~\cite{Carmichael1993QTII}. The conditioned probability density of the first trigger, resetting the charge generation process, is given as a function of the null-measurement record $|\overline{\psi}_{\rm REC,\, NULL}(t)\rangle$ as~\citep{CarmichaelBook2}
\begin{equation}\label{eq:p1}
p_1(t)=2\kappa r\frac{\langle \overline{\psi}_{\rm REC,\, NULL}(t)| a^{\dagger}a|  \overline{\psi}_{\rm REC,\, NULL}(t)\rangle}{\langle \overline{\psi}_{\rm REC,\, NULL}(t)|  \overline{\psi}_{\rm REC,\, NULL}(t)\rangle}=(2\kappa r) A^2(t) \frac{\cosh [\phi(\theta,t)]-\cos[\phi_0+\phi(\theta-\pi/2,t)]e^{-2A^2(t)}}{\cosh[\phi(\theta,t)]+\cos[\phi_0 +\phi(\theta-\pi/2,t)]e^{-2A^2(t)}},
\end{equation}
where $A(t)\equiv A e^{-\kappa t}$ and $\phi(\theta,t)\equiv 2Q_{\theta;\,1}(t)A\sqrt{1-r}\cos\theta$. In a Monte Carlo procedure without a Hilbert space, the quantity $p_1(t)\,dt$ is compared against a random number $R$ uniformly distributed between $0$ and $1$, to decide whether a jump (APD ``clock'') occurs. This is done for Fig.~\ref{fig:FIG3}. If $p_1(t)\,dt > R$, the system state is updated to
\begin{equation}
 |\psi_{\rm REC, 1}(t)\rangle=\frac{a |\overline{\psi}_{\rm REC,\, NULL}(t)\rangle}{\sqrt{\langle \overline{\psi}_{\rm REC,\, NULL}(t)|\overline{\psi}_{\rm REC,\, NULL}(t)\rangle}}.
\end{equation}

We can proceed to determine the probability density of a second reset at $t_2$ on the condition that the APD has registered the first photon ``click'' at time $t=t_1$. For $t_1<t<t_2$ we find
\begin{equation}\label{eq:p2}
p_2(t;t_1)=2\kappa r\frac{\langle \overline{\psi}_{\rm REC, 1}(t)| a^{\dagger}a|  \overline{\psi}_{\rm REC, 1}(t)\rangle}{\langle \overline{\psi}_{\rm REC, 1}(t)|\overline{\psi}_{\rm REC, 1}(t)\rangle}=(2\kappa r) A^2(t)\frac{\cosh[\varphi(\theta,t;t_1)]+\cos[\phi_0 +\varphi(\theta-\pi/2,t;t_1)]e^{-2A^2(t)}}{\cosh[\varphi(\theta-\pi/2,t;t_1)]-\cos[\phi_0+\varphi(\theta-\pi/2,t;t_1)]e^{-2A^2(t)}},
\end{equation}
where now $\varphi(\theta,t;t_1)\equiv [2Q_{\theta;\,2}(t)A(t_1)+2Q_{\theta;\,1}(t_1)A]\sqrt{1-r}\cos\theta$. Once again, comparing $p_2(t;t_1)\,dt$ against $R$ decides for the second photon emission. Here $Q_{\theta;\,2}(t)$ is the cumulative charge produced at the BHD after the first photon triggers a fresh sample making of the photocurrent. It satisfies Eq.~\eqref{eq:dQHCond}, with the potential~\eqref{eq:potential} evaluated for the updated initial state $|\psi_{\rm REC, 1}(t)\rangle$. For $\theta=\pi/2$, the term $2Q_{\theta;\,1}(t_1)A$ incorporates the effect of phase diffusion marked by the first APD photodetection event. The presence of such marker is also imprinted on the sign alternation between the numerator and denominator in Eqs.~\eqref{eq:p1} and~\eqref{eq:p2}. Phase diffusion is responsible for the dephasing of an ensemble of realizations, annihilating the interference fringes over a decoherence time.    

\subsection{Numerical generation of individual realizations in a Hilbert space}
\label{App:B2}

Finally, independently of the expressions derived in the App.~\ref{App:B1}, we implement a Monte Carlo algorithm which propagates the pure system state $|\psi_{\rm REC}(t)\rangle$ [with $\rho_{\rm REC}=|\psi_{\rm REC}(t)\rangle \langle\psi_{\rm REC}(t)|$] forward in time with a step of size $\Delta t$, in a Fock-state basis truncated at a set photon level $\sim 2A^2$ upon ensuring convergence. Results are depicted in Fig.~\ref{fig:FIG2}. The basic steps of the numerical procedure follow below~\cite{Reiner2001,Carmichael1993QTI}:

{\bf 1.} The initial state for the cavity is the normalized cat state~\eqref{eq:incat}. 

{\bf 2.} The probability for photon trigger ``click'' is calculated as
\begin{equation}
p(t)=2\kappa r\frac{\langle \overline{\psi}_{\rm REC}(t)| a^{\dagger}a|  \overline{\psi}_{\rm REC}(t)\rangle}{\langle \overline{\psi}_{\rm REC}(t)|  \overline{\psi}_{\rm REC}(t)\rangle}\,\Delta t.
\end{equation}

{\bf 3.} Associate with the photon loss channel a uniformly distributed random number $R$ between $0$ and $1$. If $p(t)>R$, then the conditioned wavefunction collapses to
\begin{equation}
|\overline{\psi}_{\rm REC}(t+\Delta t)\rangle=\sqrt{2\kappa r}\,a|\overline{\psi}_{\rm REC}(t)\rangle.
\end{equation} 

{\bf 4.} If $p(t)<R$ then $|\overline{\psi}_{\rm REC}(t)\rangle$ is propagated through SSE~\eqref{eq:SSEsup}. The field averages are calculated as
\begin{equation}
\langle a  \rangle_{\rm REC}=\frac{\langle \overline{\psi}_{\rm REC}(t)| a|  \overline{\psi}_{\rm REC}(t)\rangle}{\langle \overline{\psi}_{\rm REC}(t)|  \overline{\psi}_{\rm REC}(t)\rangle},
\end{equation} 
along with its complex conjugate. 

{\bf 5.} Normalize the system wavefunction and repeat from step 2. 

For an ensemble of normalized pure states $|\psi_{(k);\,\rm REC}(t)\rangle$, $k=1,2,\ldots N$, generated by the above procedure, the expansion of $\rho(t)$ solving the ME~\eqref{eq:ME} is approximated as a sum over records by
\begin{equation}
\rho(t)=\frac{1}{N}\sum_{k=1}^{N}|\psi_{(k);\,\rm REC}(t)\rangle \langle \psi_{(k);\,\rm REC}(t)|=\frac{1}{N}\sum_{k=1}^{N}\rho_{(k);\,\rm REC}(t).
\end{equation}
Realizations of the cumulative charge were generated (with $r=0$) by summing over the increments $dq_{\theta}$ weighted by the decaying exponential mode profile, calculated from Eq.~\eqref{eq:dxisup}, at each time step where the system wavefunction was updated through the above Monte Carlo procedure. The steady-state values matched the histograms of Fig.~\ref{fig:FIG1} obtained from the autonomous equation~\eqref{eq:dQHCond} with the potential~\eqref{eq:potentialtr} (set $\eta=1-e^{-2\kappa t}$).

\vspace*{5mm}

{\it Data availability:} The data and codes that support the findings reported herein are openly available in {\it figshare} at the set with this \href{https://doi.org/10.17045/sthlmuni.31000072}{DOI}.

\twocolumngrid

\bibliography{bibliography_LV}

\end{document}